\documentclass[pdflatex,sn-mathphys-num]{sn-jnl}

\usepackage{graphicx}%
\usepackage{multirow}%
\usepackage{amsmath,amssymb,amsfonts}%
\usepackage{amsthm}%
\usepackage{mathrsfs}%
\usepackage[title]{appendix}%
\usepackage{xcolor}%
\usepackage{textcomp}%
\usepackage{manyfoot}%
\usepackage{booktabs}%
\usepackage{algorithm}%
\usepackage{algorithmicx}%
\usepackage{algpseudocode}%
\usepackage{listings}%

\theoremstyle{thmstyleone}%
\theoremstyle{thmstyletwo}%

\theoremstyle{thmstylethree}%

\begin{document}

\title[Article Title]{Photovoltaic creation of charged domain walls in barium titanate}


\author*[1]{\fnm{Petr} S. \sur{Bednyakov}}\email{bednyakov@fzu.cz}

\author[1]{\fnm{Petr} V. \sur{Yudin}}\email{yudin@fzu.cz}
\equalcont{These authors contributed equally to this work.}

\author[2]{\fnm{Alexander} K. \sur{Tagantsev}}\email{alexander.tagantsev@epfl.ch}
\equalcont{These authors contributed equally to this work.}

\author[1]{\fnm{Ji\v{r}\'i}\sur{Hlinka}}\email{hlinka@fzu.cz}
\equalcont{These authors contributed equally to this work.}

\affil*[1]{
\orgname{Institute of Physics of the Czech Academy of Sciences (FZU)}, \orgaddress{\street{Na Slovance 1999/2}, \city{Prague 8}, \postcode{18221},
\country{Czech Republic}}}


\affil[2]{
\orgname{Swiss Federal Institute of Technology (EPFL)
}\orgaddress{\street{}, \city{Lausanne}, \postcode{CH-1015} \state{}, \country{Switzerland}}}


\abstract{The optical control of domain structures in ferroelectrics is of great interest.
In the present work, we demonstrate the reliable creation of charged domain walls~- conductive channels in otherwise insulating barium titanate~- by a combined effect of light and electric field.
We propose a scenario for the documented process, in which the bulk photovoltaic effect plays the key role, providing charge screening at the walls.
Our scenario is supported by the results of phase-field simulations.
The results are of interest for future reconfigurable electronic and opto-electronic devices.
}




\maketitle

\section{Introduction}\label{sec1}

The scientific community has long been fascinated by the physics of ferroelectrics rich in electro-mechanical, opto-electronic, and polarization-related phenomena \cite{grinberg2013perovskite}.
Remarkable activity in the field has been devoted to establishing structure-property relationships through the so-called domain engineering \cite{nataf2020domain}.
This effort recently included optical control of ferroelectric structures \cite{schroder2012conducting,rubio2018reversible, rubio2015ferroelectric}, which holds great potential as a contactless method for implementing structural changes.

Ferroelectric charged domain walls (CDWs) stand out by drastically changing material properties, acting as reconfigurable, conductive channels embedded within an insulating matrix \cite{ye2021emergent,li2016giant,bednyakov2018physics,werner2017large,maksymovych2012tunable,sluka2013free,beccard2022nanoscale}
with applications in nanoelectronics and adaptive circuitry \cite{mcCluskey2025perspective,meier2022ferroelectric,ma2018controllable}.
Realizing this potential, however, requires a fundamental understanding of the mechanisms governing their formation and stability.
In ferroelectrics, the stability and formation of CDWs are intrinsically linked to charge compensation \cite{gureev2011head}.
The nature of this compensation depends critically on the available charge carriers and their dynamics.
Traditionally, such compensation was viewed as controlled by the conductive current supported by the mobile carriers of the material \cite{sluka2013free,bednyakov2015formation} or electron-hole pairs generated under super-bandgap illumination \cite{bednyakov2016free,sturman2017charged}.
Recently, it was pointed out \cite{liou2025phase} that the charge transport due to the bulk photovoltaic effect (BPVE) also favors CDW formation.
However, currently, the role of super-bandgap illumination in CDW formation remains unclear.

In this work, we experimentally studied BaTiO$_3$ (BTO) crystals containing a system of neutral domain walls (NDWs) under ultraviolet illumination and a DC electric field, demonstrating the reproducible conversion of such a system into a pattern of CDWs.
We propose a scenario for the documented process, in which the bulk photovoltaic effect plays the key role, providing charge screening at the walls.
Our scenario is supported by the results of phase-field simulations.
Our work provides clear experimental evidence for the important role of the BPVE in CDW formation.
On the practical side, our results demonstrate an original way of optical control of domain structures in ferroelectrics, which may be of interest for future reconfigurable electronic and opto-electronic devices.
%

\section{Results}\label{sec2}
\subsection{Experimental settings}\label{subsec2.1}

\begin{figure}[h]
\centering
\includegraphics[width=0.7\textwidth]{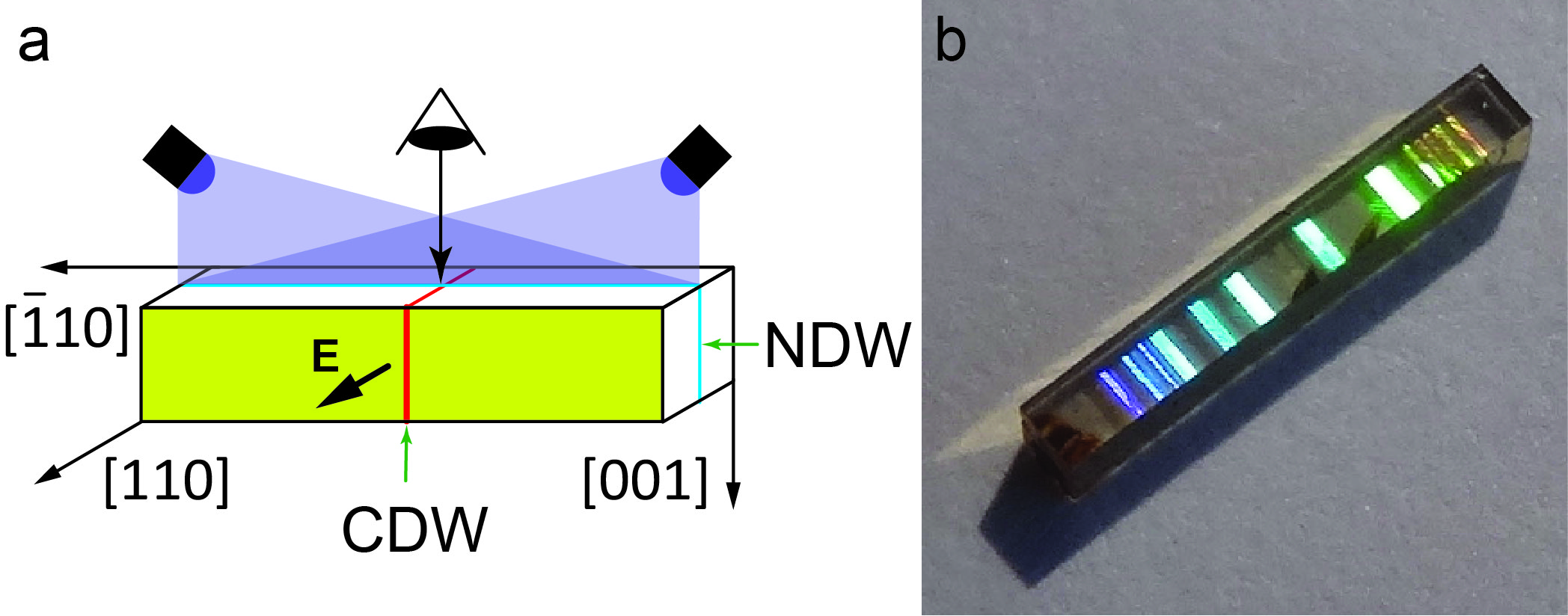}
\caption{
The geometry of the experiment (a).
The orientations of charged and neutral domain walls are marked with red and blue lines, respectively.
The sample is illuminated with UV light and observed from the [001] direction; the electric field is applied along the [110] direction.
(b) A photograph of the (001) surface of a typical sample with ferroelectric CDWs.
Domain walls appear as colored stripes due to light interference when viewed slightly off from a top-down position.
}
\label{Sample}
\end{figure}
%
\begin{wrapfigure}{r}{0.5\textwidth}
\centering
\includegraphics[width=0.5\textwidth]{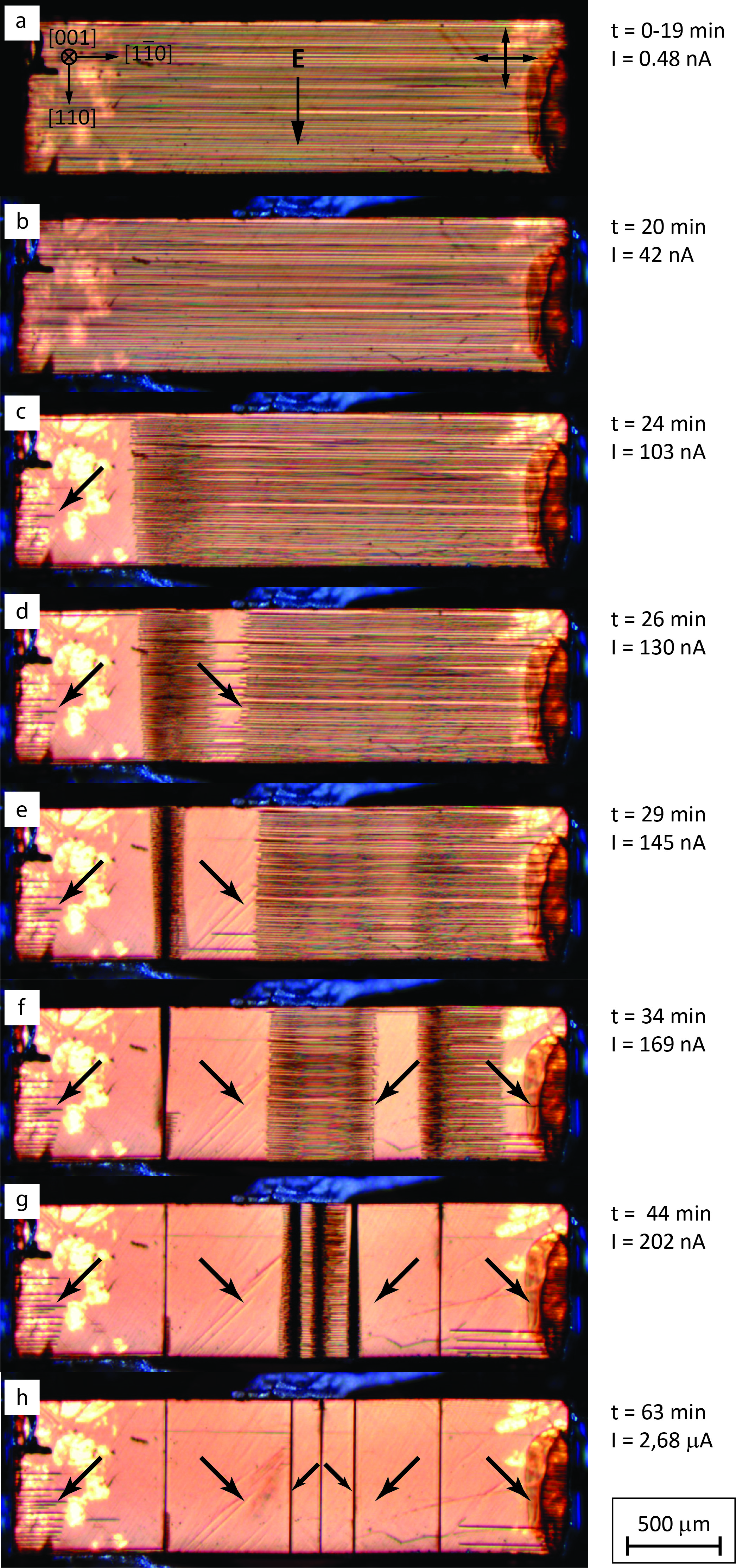}
\caption{
The (001) plane view of CDW formation.
NDWs are seen as the horizontal lines in a-f, CDWs - as the vertical lines in f-h.
(a) The domain structure under an applied voltage of 300 V, which is identical to that initial (no UV).
(b–g) The  evolution of the domain structure when the voltage is kept while the sample is UV illuminated. 
(h) Final configuration with conductive CDWs.
The arrows show the direction of the spontaneous polarization.
}
\label{PrimaryCDWs}
\end{wrapfigure}
We experimentally studied the evolution of the domain structure in BTO single crystals.
The samples had a bar-like geometry with dimensions of 2.87 × 0.65 × 0.8 mm, with the 2.87 × 0.65 mm faces, which were electroded, being parallel to the (110) crystallographic plane (hereafter, the cubic indexing is used).
The longest edge was parallel to the $[1\bar{1}0]$ direction (Fig.\,\ref{Sample}).
The samples were under a voltage applied to the electrodes, which induced the electric field on average along the [110] direction.
Under such conditions, two directions of the spontaneous polarization are favored by the field, [100] and [010].
From these polarization states, three simple domain configurations can be conceived: the single-domain configuration and two multidomain configurations: (a) consisting of NDWs with domain walls parallel to the (110) plane and (b) consisting of CDWs with domain walls parallel to the $(1\bar{1}0)$ plane.
One readily expects that the single-domain configuration must be favorable.
However, we recently showed \cite{bednyakov2024paradoxical} that, unexpectedly, in the BTO samples of the above geometry, if a large enough voltage is applied, the mainly single-domain configuration converts into the (a) multidomain pattern.
In this work, we studied the impact of the ultraviolet illumination on such a domain pattern and found that, in the presence of a sufficiently large applied voltage, the illumination readily converts it into the (b) configuration.
We illuminated the non-electroded (001) face with ultraviolet light of 365 nm wavelength and intensity of about 10 mW/cm$^2$.
The light absorption coefficient $\alpha_{abs}=5\times10^3$ m$^{-1}$ was obtained by measuring the transmitted light intensity. Thus, the light penetration depth of 0.2 mm is comparable to the thickness of the sample.
%

\subsection{CDWs formation}\label{subsubsec2.1.2}
Depending on the sample's prehistory and variations in the intensity of UV illumination, the process of CDW formation took between 10 minutes and 2 hours.
The typical evolution of the domain structure is presented in Fig.\,\ref{PrimaryCDWs}
.
Here, the images correspond to 3D domain structures, in which the cross-sections normal to the [001] direction are the same.

First, a voltage of 300 V was applied to the sample, which initially contained the aforementioned (a) domain configuration of NDWs.
This produced an electric field of about 4 kV/cm, oriented on average parallel to the [110] direction.
This field supports the sign of the polarization in all domains, and the domain structure remained unchanged (Fig.\,\ref{PrimaryCDWs}~a).

Next, keeping the applied voltage, the sample was exposed to UV illumination.
This resulted in the following evolution of the domain structure.
The planar (110) NDWs near the sample edges began to vanish or distort, producing a local imbalance in domain fractions as one polarization state expanded at the expense of the other (Fig.\,\ref{PrimaryCDWs}~c–d).
Subsequently, zigzag patterns form, separated by monodomain regions (Fig.\,\ref{PrimaryCDWs}~d–e).
Ultimately, the zigzags shrink to form planar $(1\bar{1}0)$ CDWs (Fig.\,\ref{PrimaryCDWs}~d–h).
We found that the growth rate of CDWs increases with increasing electric field strength.
\begin{figure}[t]
\centering
\includegraphics[]{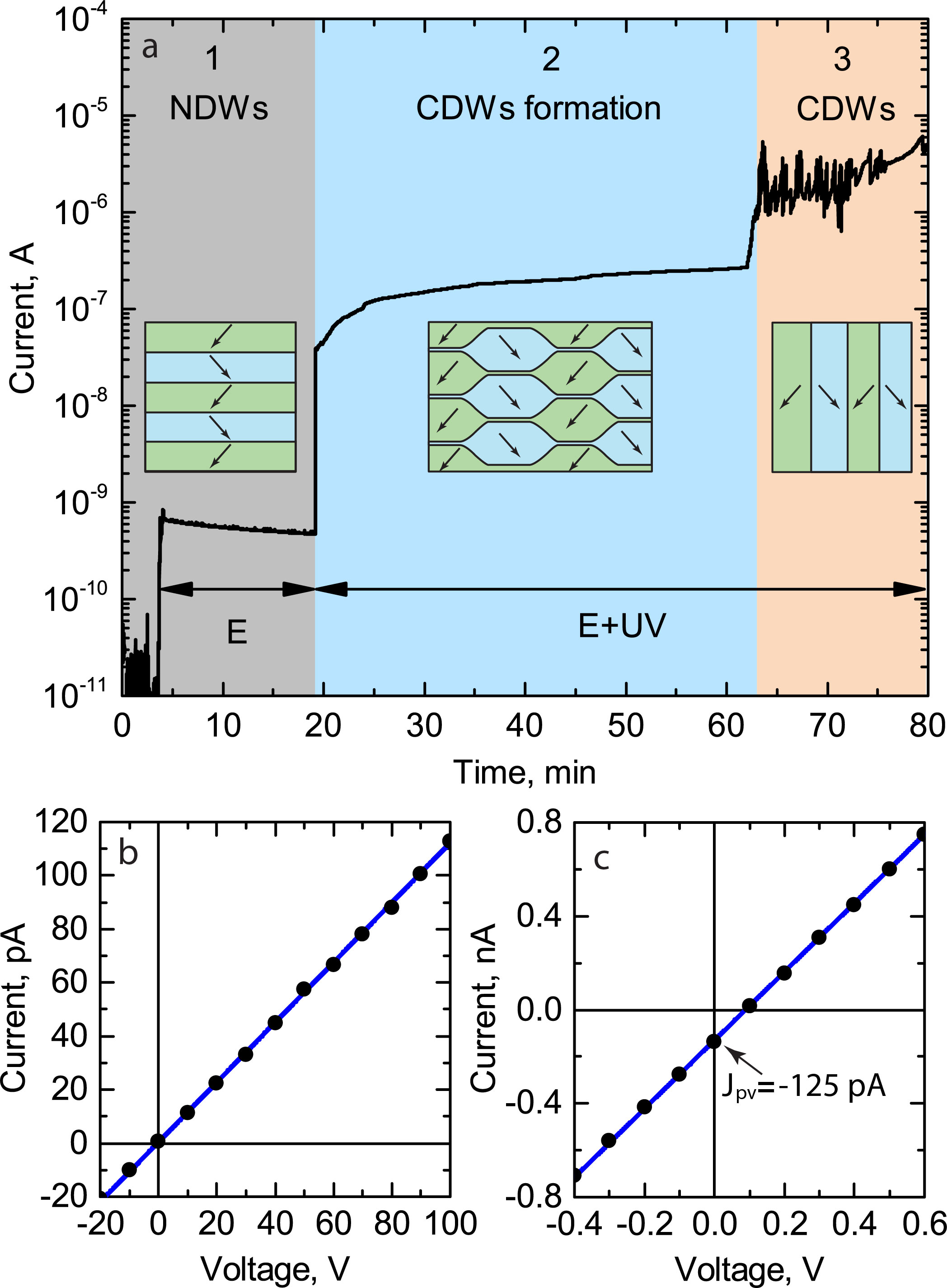}
\caption{
The time evolution of current during the NDW-to-CDW transformation, with the schematics of the domain structures.
(a) Region~1: The voltage is applied and no illumination is present.
No change in the domain structure is observed.
Region~2: The voltage is kept.
Under UV illumination, bending of NDWs and the formation of zigzag patterns are observed.
Region~3: The system of CDWs is formed.
I–V curves for the sample with numerous NDWs without UV illumination (b) and with it (c).
}
\label{Current}
\end{figure}

We also tested the impact of the illumination in the absence of the electric field.
We found that, without the field, the illumination itself leads to the disappearance of some domains separated by NDWs, but CDWs were not created, at least during a reasonable observation time.
At the same time, once the zigzag was already formed (Fig.\,\ref{PrimaryCDWs}~e), the illumination can, in principle, drive the process to completion; however, it proceeds extremely slowly, and full CDW formation was never achieved within the experimental time frame.

It is worth noting two more features of the resulting CDW pattern: (i) an odd number of CDWs always appears, (ii) in the resulting CDW pattern, in the side domains, the polarization points toward the side surfaces.

To obtain more information on what is happening in the sample, we performed current measurements in the applied voltage circuit.
In the absence of UV illumination, the sample exhibited a nearly ohmic I–V response (Fig.\,\ref{Current}~b).
At zero voltage, only current fluctuations of about $1\times10^{-12}$~A were observed, while with an applied voltage of 300 V, a steady current of approximately $5\times10^{-10}$~A was produced (Fig.\,\ref{Current}~a, Region~1).

Upon UV illumination, the generation of free carriers caused an abrupt two-order-of-magnitude increase in current (Fig.\,\ref{Current}~a).
Over the subsequent 40 minutes, the current continued to rise gradually by about half an order of magnitude, following the progression of CDW formation (Fig.\,\ref{Current}~a, Region~2).
Once conductive channels corresponding to $(1\bar{1}0)$-oriented CDWs formed in the sample (Fig.\,\ref{PrimaryCDWs}~h), the current dramatically increased by an additional two orders of magnitude, reaching $\sim10^{-5}$ A (Fig.\,\ref{Current}~a, Region~3).
We consider this stage to be the final step in the formation of CDWs.
The current behavior in Region 3 is notably non-monotonic, indicating the presence of dynamic processes occurring within the crystal containing CDWs.
These processes are important and merit further investigation, but they fall outside the scope of this paper.

We also performed low-voltage current measurements both without (Fig.\,\ref{Current}~b) and under (Fig.\,\ref{Current}~c) illumination and found the nearly ohmic behavior.
Under illumination and at zero voltage, we observed (Fig.\,\ref{Current}~c) a current related to BPVE, $J_\text{pv}=-125$~pA, flowing against the direction of polarization (Fig.\,\ref{Current}~c).
The corresponding value of the photovoltaic coefficient was evaluated as $\frac{J_{PV}}{I A_{\text{ill}}} \approx 10^{-6}$ C/J, where $A_{\text{ill}}$ and $I$ are the area of the side face of the sample that is illuminated and the light intensity, respectively.
This estimate agrees with literature data \cite{zenkevich2014giant,shafir2023ultrahigh}.

Our principal observation is that the UV illumination destabilizes the periodic configuration of NDWs, while the additional presence of a sufficiently large electric field promotes the formation of a system of CDWs.
%

\subsection{Interpretation of the  domain kinetics and modeling}\label{subsec2.2}
\begin{wrapfigure}{r}{0.5\textwidth}
\centering
\includegraphics[width = 0.5\textwidth]{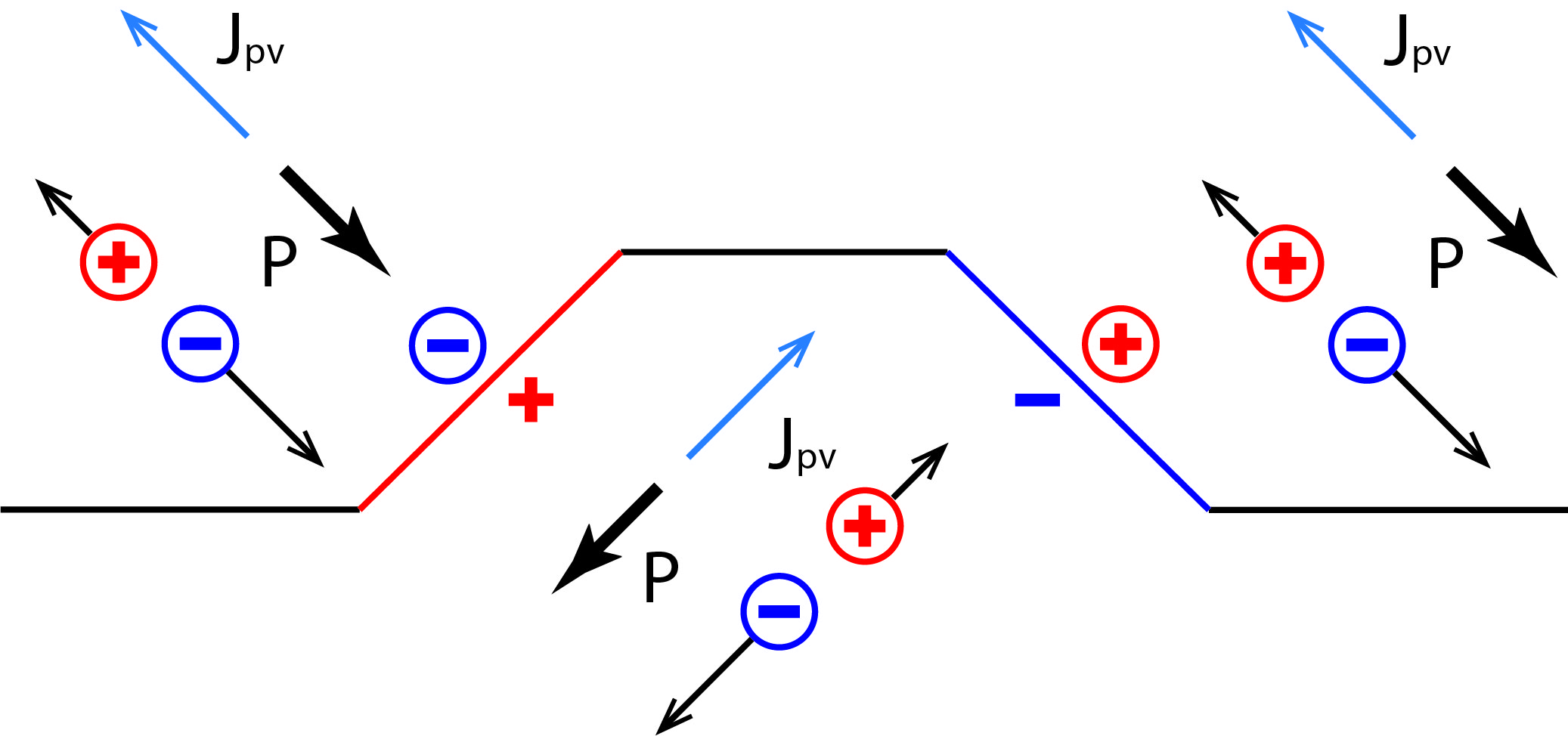}
\caption{The photovoltaic mechanism for the destabilization of NDWs and the growth of CDWs.}
\label{hunp}
\end{wrapfigure}
Our experimental data imply that UV illumination promotes the formation of CDWs.
It is currently understood that such a formation requires the screening of the bound charge at the walls with free carriers \cite{sluka2013free,sturman2015quantum,sturman2017charged}.
Under the UV illumination, one can distinguish two channels for the charge transport needed for such a screening: (i) the conductive current driven by the electrostatic potential difference and (ii) the photovoltaic current, which does not require any potential difference.
An important question to be addressed is which of the channels plays the main role in the phenomenon we experimentally documented in this paper.
We propose the following scenario, where the photovoltaic current plays the decisive role.

\begin{figure}
\centering
\includegraphics[width=\textwidth]{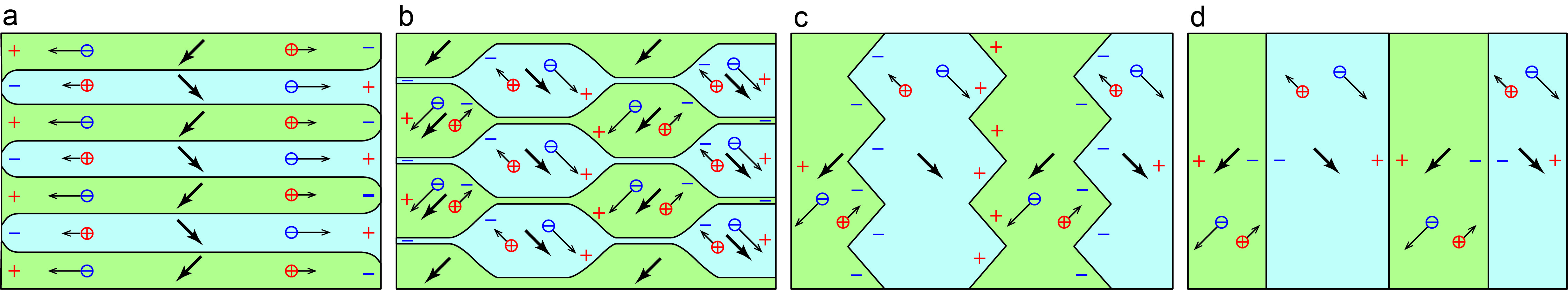}
\caption{Schematic for the stages of CDW growth: (a) appearance of original domain inclinations due to the asymmetry of the screening of sample edges with positive and negative free carriers; (b) growth of kinks due to the photovoltaic effect and their alignment into vertical columns; (c) formation of zigzag CDWs; (d) the final structure containing vertical CDWs.
The view is taken in the same direction as in Fig.\,\ref{PrimaryCDWs}.
\label{Schematic}}
\end{figure}

It is known that, in BTO, the photovoltaic contribution to the electric current is antiparallel to the spontaneous polarization \cite{inoue2015giant}, which is consistent with our current measurements.
This photovoltaic current, propagating opposite to the polarization, promotes the accumulation of free charge at the boundaries between domains that differ in the normal components of the polarization.
In the initial configuration of NDWs, the above components are equal, and no free-charge accumulation will take place.
However, if a small hump protrudes from the wall (Fig.\,\ref{hunp}), it starts accumulating free charge of the sign opposite to that of the bound charge.
Such an accumulation, in general, will not stop when the bound charge is compensated, such that the latter may be overcompensated.
Once it is overcompensated, since the bound charge per unit area of the wall increases with increasing angle between the sides of the hump and the original plane of the NDW, this angle will increase, trying to keep the neutrality of the wall.
This way, the hump will grow.
This mechanism may destabilize the original pattern of NDWs.
In the further evolution of the domain pattern of the sample, the photovoltaic current flowing against the polarization, as an effective screening mechanism, will also promote the formation of CDWs.
We suggest that the NDWs can be easily destabilized at the sides of the sample, where the initial deviation of the NDWs from the neutral orientation is readily expected, since the fractions of the two domain states at the side faces may be unequal, as shown in Fig.\,\ref{Schematic}~a.
The reason for this asymmetry is as follows.
In BTO, the charge transport at ambient conditions is provided by electrons \cite{erhart2008modeling}.
We assume that the photovoltaic current is also dominated by electrons.
This implies more efficient screening of the positive polarization bound charge on the side faces of the sample, promoting the larger width of the corresponding domain at these faces, as shown in Fig.\,\ref{Schematic}~a.
This effect also explains our observation that, in the resulting domain pattern, the polarization points toward the side faces, Fig.\,\ref{PrimaryCDWs}.

Keeping in mind the above arguments, we conceive a scenario for the evolution of the domain pattern in the system, that is schematically depicted in Fig.\,\ref{Schematic}.

There is also an indirect indication of the key role of the BPVE.
The time $\tau$ needed for CDW formation, which we reported, is huge on typical solid-state time scales.
However, assuming that the BPVE controls the story, it can be readily rationalized: if it is this effect that controls the screening in the final CDW configuration, this time can be roughly evaluated as the time needed to supply the free charge for the screening of the bound charge on CDWs, which reads
\begin{equation}
\label{Timescale}
\tau \approx \frac{\sqrt{2}P_S}{\Delta j_\text{pv}}
\end{equation} 
where $\Delta j_\text{pv}$ is the difference between the densities of the photovoltaic currents at the two sides of the wall.
During the evolution of the domain pattern, it changes from very small values to $2 J_\text{pv}/A_{\text{eff}}$ where  $J_\text{pv}=-125$~pA is the photovoltaic current we measured and $A_{\text{eff}}\approx 0.6\times10^{-6}$ m$^2$ is the area of the efficiently working part of the electrode.
Using these values, we find $\tau \approx 900$ s as a lower estimate for the time needed for the photovoltaic current to provide the screening in the resultant CDW pattern.
This value fairly correlates with the time scales of the NDWs-to-CDWs conversion we reported.

To support the above scenario, we performed phase-field simulations of the system under conditions of partial similarity because quantitative modeling of the domain evolution was too challenging.
The reason is that the domain transformation involves processes whose characteristic times span about 15 orders of magnitude, from $10^{-12}$ s for the relaxation of the polarization to 1000\,s for the formation of the final domain pattern.
Also the spatial scales span 6 orders of magnitude from 1\,nm for an individual DW to 1\,mm  of the sample size.
%
%
%
Thus, the quantitative description in question would require multiscale modeling.
The numerical equation solver available to us was not suitable for such modeling since, for realistic computation times, with the material parameters of BTO and the actual UV light intensities, it provided a description of the evolution of the system only up to $10^{-10}$\,s.
Nevertheless, we decided to use this solver to obtain some qualitative arguments supporting our supposition that it is the BPVE that makes the documented transformation of the domain pattern possible.
Our approach was as follows: we considered the same computation scheme and looked for a set of modified input parameters (material parameters and the UV light intensity), which accelerate the domain pattern transformation such that it can be simulated during a reasonable computation time, and we compared the results of the simulations with the experiment and our scenario.
\begin{wrapfigure}{r}{0.5\textwidth}
\centering
\includegraphics[width=0.5\textwidth]{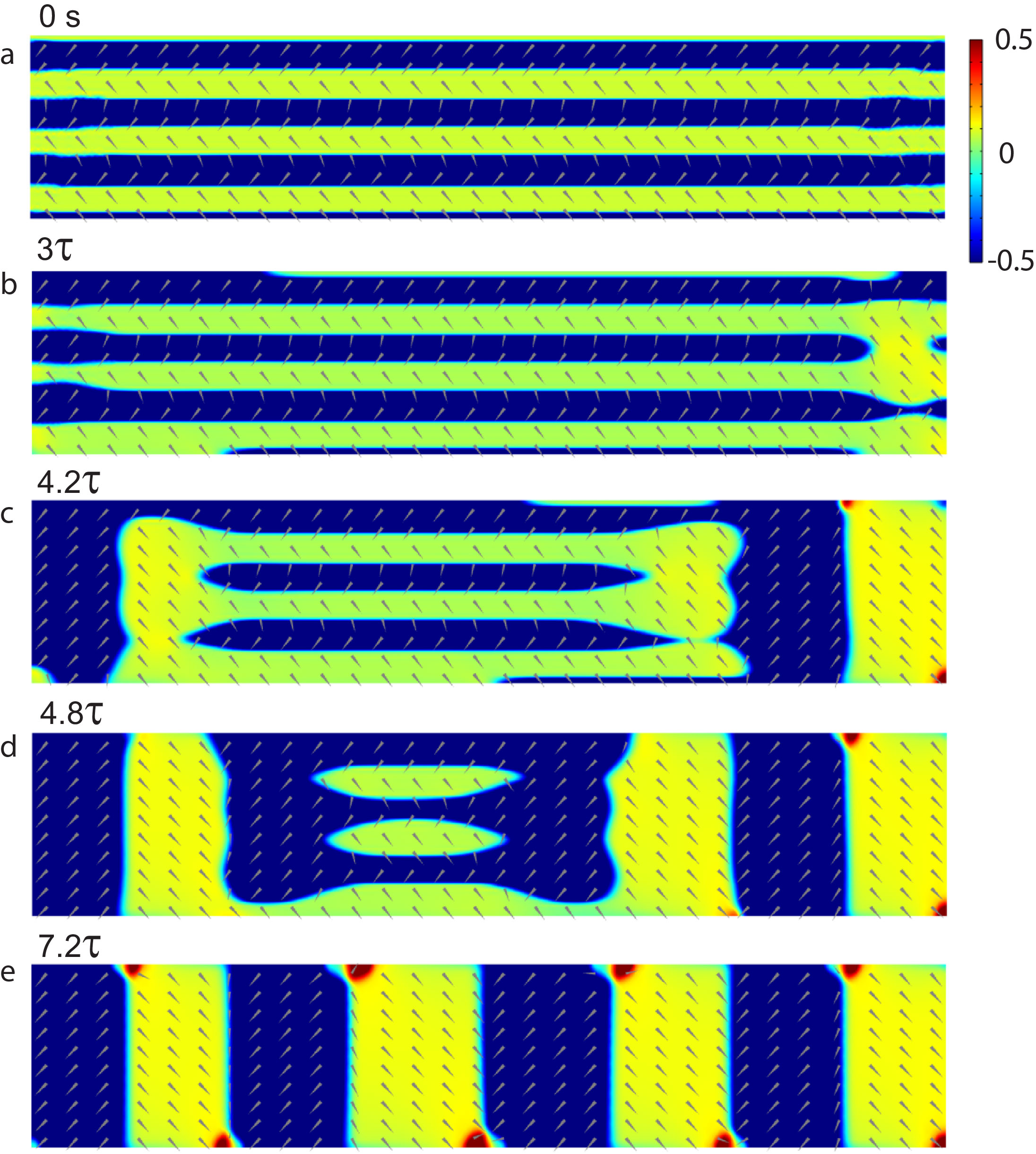}
\caption{Phase-field simulations of the domain pattern evolution.
(a) The original system of horizontal neutral domain walls; (b–d) the intermediate stages of the transformation; (e) the final system of vertical charged domain walls.
Arrows show polarization; the color map is introduced for the value $1.5 P_x + P_y$.
Units of characterization time $\tau $ are used for the Eq. (\ref{Timescale}) with model parameters.
The view is taken in the same direction as in Fig.\,\ref{PrimaryCDWs}.
\label{scenario}}
\end{wrapfigure}
We managed to find such a set of modified input parameters. It is given in Table \ref{tab:par} together with the original set.
Aiming to reproduce only some qualitative features of the phenomenon, we permit ourselves an additional simplification: we consider homogeneous, through-the-sample generation of photo-carriers.
This allows us to use 2D simulations for the description of the domain transformation.

In our simulations performed using the set of modified input parameters, the details of which are given in Sect.~\ref{secsimul} of Methods, we started from the periodic pattern of NDWs identical to that which was used as the initial state of the experiment (Fig.\,\ref{PrimaryCDWs}).
We kept it under a DC voltage and super-bandgap illumination.
Next, we explored the contribution of BPVE to the charge transport in the system.
First, we turned the BPVE off such that the illumination manifested itself only in the generation of photo-carriers, and we found no modification of the initial system of NDWs.
Next, we turned the BPVE on in two ways: (i) with the photovoltaic current flowing against the spontaneous polarization and (ii) with the photovoltaic current flowing along the spontaneous polarization.
On the lines of our scenario, the effect is expected only in the first case.
Indeed, we obtained the NDWs-to-CDWs conversion only in the first case, while, in the second case, the initial system of NDWs was not affected by turning the BPVE on.

The results of the simulations with the photovoltaic current flowing against the spontaneous polarization are presented in Fig.\,\ref{scenario} showing a qualitative agreement with the experimental observations.
Fig.\,\ref{scenario}~a displays the set of planar NDWs used as the initial condition for the simulation. 
Fig.\,\ref{scenario}~b illustrates an early stage, where domains with polarization pointing outside the sample begin to dominate near the sample edges.
Fig.\,\ref{scenario}~c and~d present the subsequent stages, in which CDWs first appear near the edges and later emerge in the central region.
Finally, Fig.\,\ref{scenario}~e shows the system of vertical CDWs to which the simulation converges.

As is seen from Fig.\,\ref{scenario}, the simulated polarization at the sides always points towards the adjacent faces of the sample, as in the experiment.
Above, this feature of the system was related to some asymmetry in the polarization screening at the sides of the sample.
The simulations suggest that it originates from the asymmetry of photovoltaic currents from electrons and holes.
Specifically, the domain patterns shown in Fig.\,\ref{scenario} were obtained when the photovoltaic current of electrons was set larger than that of the holes.
Once we swapped this relation, the orientation of the polarization was swapped as well.

At our experiments, for the efficient NDW-to-CDW conversion, the common action of the applied voltage and illumination was needed.
We repeated the above simulation, but in the absence of the applied voltage.
We found some evolution of the initial domain pattern (Supplemetary Fig.\,\ref{NoField}), however it did not end in the regular system of CDWs observed in the experiment.

All in all, we provided a scenario for the NDWs-to-CDWs conversion experimentally documented using qualitative arguments.
We also performed the numerical simulation of our system, however, with strongly modified values of the input parameters.
The fact that it reproduces a number of qualitative features of the experimental observations and proposed scenario provides additional support for the latter.
%

\section{Conclusion}\label{sec13}

We documented the transformation of the system of neutral domain walls into that of charged domain walls in a BTO sample exposed to UV illumination and DC electric field.
Our experiments and phase-field simulations strongly suggest that such a transformation is conditioned by the interplay of the initial domain structure, the BPVE, and an applied electric field.
This behaviour stems from persistent BPVE-driven electron pumping, which induces CDW overcompensation and drives their evolution toward a strongly charged planar state.
Our findings not only clarify the origin of CDW formation but also establish a pathway for exploiting this phenomenon in photovoltaic conversion applications.

\section{Methods}\label{sec11}
\subsection{Experimental technique}
$\langle 110 \rangle$-oriented samples of BaTiO$_3$ monocrystal were obtained from Electro-Optics Technology GmbH.
These crystals were grown using the top-seeded solution growth (TSSG) technique and had a bar-like shape with dimensions of 2.87 × 0.65 × 0.8 mm, with the longest edge aligned along the $[1\bar{1}0]$ direction (Fig.\,\ref{Sample}~a).
The (001) crystal planes were polished to a surface quality of 1\,\textmu m, allowing for domain structure observation along the [001] direction.

To apply an electric field, the (110) surfaces were coated with a ~100 nm thick gold electrodes.
The electric field was applied on average along the [110] direction using a Keithley 6517B electrometer equipped with an integrated 1\,kV DC power supply.
Electrical contacts to the electrode surfaces were made using high-temperature silver paste.

Free charge carriers were generated through super-bandgap illumination using a 365\,nm ultraviolet (UV) light source with a luminous flux of approximately 5\,W, based on an OSRAM LZ4-V4UV0R light-emitting diode.
To estimate the intensity of effectively delivered UV illumination to the sample and the penetration depth we used a PM100A Compact Power Meter Console with a S120VC Photodiode Power Sensor from THORLABS, allowing measurement of illumination power of 200 nm–1100 nm wavelength sources.
UV illumination was delivered to the sample via an optical fiber, while the sample was placed in a Linkam LNP95 cell under an optical microscope.
The measured UV intensity on the sample surface was approximately ~10$\pm 2$\,mW/cm$^2$.
The light absorption coefficient $\alpha_{abs}=5\times10^3$m$^{-1}$ was established by a careful measurement of the transmitted light intensity. Thus, the light penetration depth through the non-electroded surface of 0.2~mm is comparable with the size of the sample.

Samples with various domain structures were analyzed using a Leica DM2700M polarizing microscope in both transmission and reflection modes.
The process of CDW formation was documented using Leica LAS X software.
All experimental procedures were automated and controlled within the LabVIEW environment.

\subsection{NDWs engineering technique}

\begin{figure}[h]
\centering
\includegraphics[width=0.6\textwidth]{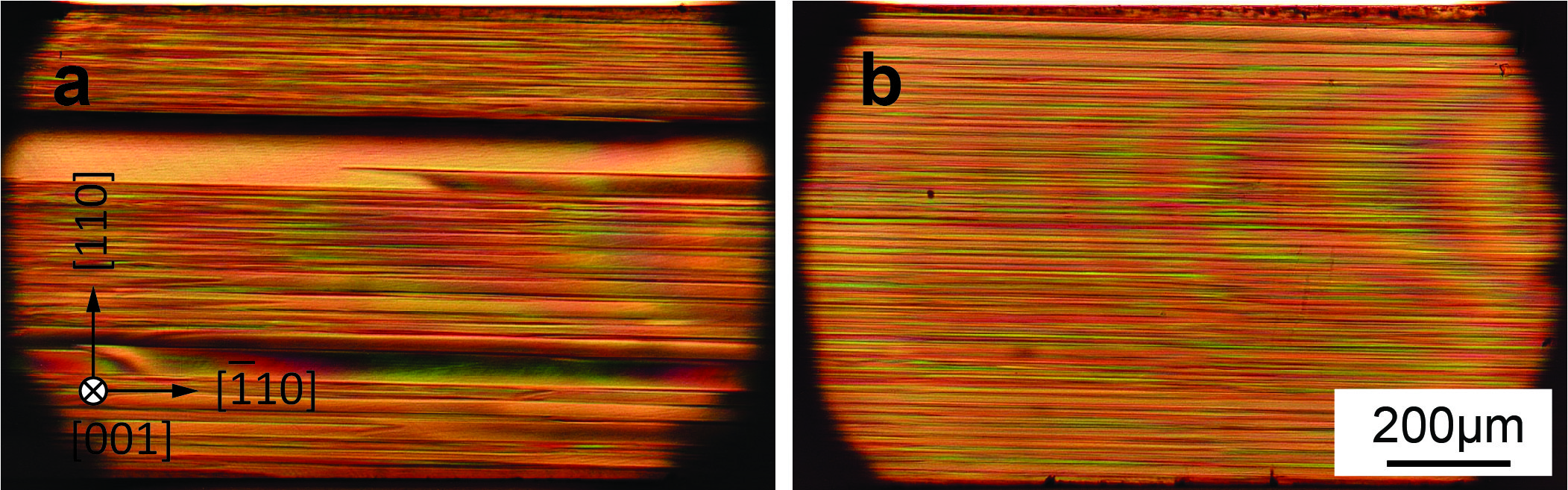}
\caption{
The specific domain structure required for CDW formation was obtained using two different methods: (a) poling at room temperature and (b) poling through the phase transition from the cubic to the tetragonal phase.
}
\label{Samples}
\end{figure}

The first method, described in detail in Ref. \cite{bednyakov2024paradoxical}, involves the formation of domain wedges in the (110) plane, separated by (110) NDWs when an external electric field is applied along the [110] direction.
The critical field required to induce this process was experimentally determined and analytically calculated to be approximately 2\,kV/cm.
Following this poling procedure, a regular domain structure with (110) domain walls spaced at approximately 3-10\,\textmu m is formed (Fig.\,\ref{Samples} a).
The second method involves poling the crystal during the phase transition from the cubic to the tetragonal phase.
This approach results in the formation of a domain structure with a constant polarization projection along the applied electric field direction.
In contrast to the first method, the required electric field for this process is significantly lower, even below 1~kV/cm.
The outcome is a random (110) laminar domain structure, where domains with (100) and (010) polarization states are separated by (110) NDWs spaced at approximately 3\,\textmu m (Fig.\,\ref{Samples}~b).
%

\subsection{Simulations}
\label{secsimul}
\subsubsection{Governing equations}
The model equations were obtained using the Lagrange principle from the enthalpy density \cite{sluka2013free}:

\begin{eqnarray}\label{LGD}
H[\{P_i,P_{i,j},e_{ij},E_i\}]=H_{\text{bulk}}^{(e)}+H_{\text{ela}}+H_{\text{es}}+H_{\text{wall}}+H_{\text{ele}},
\end{eqnarray}
where $P_i$ denotes the ferroelectric part of the polarization, $P_{i,j}$ its derivatives (the subscript '${,i}$' represents the operator of spatial derivatives $\partial/\partial x_i$), $E_i=-\varphi_{,i}$ is the electric field, $\varphi$ is the electric potential, and $e_{ij}=1/2(u_{i,j}+u_{j,i})$ is the elastic strain, where $u_i$ is the displacement vector.

The bulk enthalpy density
{\small \begin{eqnarray}\label{fbulk}
\nonumber \lefteqn{H_{\text{bulk}}^{(e)}[\{P_i\}]=}\\
\nonumber  &&\alpha_{1}\displaystyle\sum_i P_i^2+\alpha_{11}^{(e)}\displaystyle\sum_iP_i^4+\alpha_{12}^{(e)}	 \displaystyle\sum_{i>j}P_i^2 P_j^2+\alpha_{111}	\displaystyle\sum_{i}P_i^6\\
 &&+\alpha_{112}	\displaystyle\sum_{i>j}(P_i^4 P_j^2+P_j^4 P_i^2)+\alpha_{123}	 \displaystyle\prod_{i} P_i^2
\end{eqnarray}}
is expressed for a zero strain (moduli which are different from the commonly introduced stress-free ones are  highlighted with superscript (e))  as a sixth-order polynomial expansion \cite{hlinka2009piezoelectric}, where $\alpha_i,
\alpha_{ij}^{(e)}, \alpha_{ijk}$ are parameters fitted to the single crystal properties (Table \ref{tab:par}).
The remaining contributions represent bilinear forms of densities of the elastic energy
$H_{\text{ela}}[\{e_{ij}\}]=1/2 c_{ijkl} e_{ij} e_{kl}$, where $c_{ijkl}$ is the elastic stiffness,
electrostriction enthalpy $H_{\text{es}}[\{P_i,e_{ij}\}]=-q_{ijkl} e_{ij} P_k P_l$, where $q_{ijkl}$ are the electrostriction coefficients, gradient energy $H_{\text{wall}}[\{P_{i,j}\}]=1/2 G_{ijkl} P_{i,j} P_{k,l}$, where $G_{ijkl}$ are the gradient energy coefficients, and electrostatic enthalpy $H_{\text{ele}}[\{P_i,E_i\}]=-\varepsilon_0 \varepsilon_B E_i^2/2  -E_i P_i$, where
$\varepsilon_0$ and $\varepsilon_B$ are permittivity of vacuum and relative background permittivity, respectively.

The set of field equations governing the kinetics of ferroelectrics is as follows:
\begin{eqnarray}\label{eqs}
\label{mech}\left(\frac{\partial H}{\partial e_{ij}}\right)_{,j}&=&0,\\
\label{GD}\frac{1}{\Gamma}\frac{\partial P_{i}}{\partial t}-\left(\frac{\partial H}{\partial P_{i,j}}\right)_{,j}&=&-\frac{\partial H}{\partial P_{i}},\\
\label{diel}\left(\frac{\partial H}{\partial E_{i}}\right)_{,i}&=&e(p-n+N_{\bullet}).
\end{eqnarray}

Equation (\ref{mech}) defines the mechanical equilibrium, with inertia neglected.
Equation (\ref{GD}) is the time dependent Landau-Ginzburg-Devonshire equation \cite{Semenovskaya1998} that governs the spatio-temporal
evolution of spontaneous polarization with kinetics given by the coefficient $\Gamma$.
The Poisson Eq. (\ref{diel}) represents Gauss's law for charge and electric field in a dielectric that includes a nonzero concentration of free electrons $n$, holes $p$, and ionized donors (filled traps) $N_{\bullet}$. 
Donors are considered static and  present in the material with uniform and constant concentration $N=N_{\circ}+N_{\bullet}$.
This simple model also accounts for donors with valency different from 1 through appropriate renormalization.
The balance equations for the free charge carriers, accounting for light absorption, take the following form \cite{sturman2017charged}:
\begin{eqnarray}\label{cbal}
\label{ebal}\frac{\partial n}{\partial t}-\frac{1}{e} j^{(n)}_{i,i}=g-\gamma_nnN_{\circ}-\gamma np,\\
\label{nbal}\frac{\partial p}{\partial t}+\frac{1}{e} j^{(p)}_{i,i}= g-\gamma_ppN_{\bullet}-\gamma np,\\
\label{dbal}\frac{\partial N_{\bullet}}{\partial t}=\gamma_nnN_{\circ}-\gamma_ppN_{\bullet}-\gamma np.
\end{eqnarray}
Here $\gamma_{n}$, $\gamma_{p}$ and $\gamma$ are recombination constants characterizing the linear and quadratic recombination, $g$ is the rate of light-induced transitions,
\begin{eqnarray}\label{currents}
\label{jel}j^{(n)}_i=e \mu_n n E_i +e D_n n_{,i}+e B_n n P_i\\
\label{jp}j^{(p)}_i=e \mu_p p E_i -e D_p p_{,i}+e B_p p P_i.
\end{eqnarray}
are the electron and hole current densities, $\mu_{n}$, $\mu_{p}$,  $D_{n}$ and $D_{p}$ are the corresponding mobilities and diffusion coefficients, $B_{n}$ and $B_{p}$ are the terms simply describing the bulk photovoltaic effect (BPVE), contributed by  electrons and holes respectively. 
Their relation to the classically introduced photovoltaic tensor is explained in supplementary section \ref{tensorbeta}.
Thermal excitation is neglected so that the  excitation rate $g$ can be expressed by the light absorption coefficient $\alpha_{abs}$:
\begin{equation}
\label{Absorption}
g=\frac{I}{\hbar \omega w}.
\end{equation}

\subsubsection{Input parameters}
Table \ref{tab:par} gives the parameters characterizing the system considered and the input parameters of the modeling.
The value of the photovoltaic coefficients $B_e$ and $B_h$ marked in this Table with (*) was obtained under the assumption that both the photocurrent and the photovoltaic current are supported by one kind of carriers.
Their concentration was estimated as $n_{\text{exp}}\approx\frac{J h}{A_{\text{eff}} U \mu e}\approx 1.35\times10^{17}$m$^{-3}$, where $e=1.6\cdot10^{-19}$C is the elementary charge, $J=5\times10^{-8}$~A is the measured photocurrent, $U=300$ V is the applied voltage, $\mu\approx 10^{-5}$ m$^2$/(V$\cdot$s) \cite{yoo2004electronic} is the mobility of the carriers, and $h=0.8$~mm is the sample height.
Here, $A_{\text{eff}}\approx 0.6\times10^{-6}$ m$^2$ is the area of the efficiently working part of the electrode, which is about 1/3 of the electrode area.
Based on this current value and on the carrier concentration, we calculated the photovoltaic coefficient: $B_n=\frac{\sqrt{2}J_{PV}}{eA_{\text{eff}}n_{\text{exp}}P_S}=-0.053$ m$^3$/(C$\cdot$s), where $P_S=0.26$ C/m$^2$ \cite{jona1993ferroelectric} is the spontaneous polarization of BTO.

The mobilities and diffusion coefficients of free carriers given in the table are related by the Einstein relation.
This relation is not valid for the generated gas of carriers, which is expected to occur in the cores of CDWs \cite{sturman2017charged}.
However, it can be used for the description of charge transport outside the cores of CDWs, in which we are interested.

\begin{table}[h!]
\caption{The values of the parameters characterizing the system considered (material parameters for BTO, photon energy, and the intensity of UV light) and those used in the modeling are given in the third column.
If they are altered in simulations, the values are shown in a separate column.
* - see comments in the text.
** - values consistent with photocurrent measurements.
*** - measured from the intensities of incident and transmitted light.
}\label{tab1}%
\begin{tabular}{@{}lllll@{}}
\toprule
Parameter & Unit & Experimental value &  Modeling value & Description \\
\midrule
 \hline
  $\alpha_1$ & $Jm/C^2$ & $(T-381) 3.34 \times10^{5}$ & &  \\
  $\alpha_{11}^{(e)}$ & $Jm^5/C^4$ & $(T-393) 4.69 \times10^{6}+6.15\times10^8$  & & \\
  $\alpha_{12}^{(e)}$ & $Jm^5/C^4$ & $-3.67\times10^8$ & &      \cite{hlinka2009piezoelectric} \\
  $\alpha_{111}$ & $Jm^9/C^6$ & $-(T-393) 5.52 \times10^{7}+2.76\times10^9$ & &\\
  $\alpha_{112}$ & $Jm^9/C^6$  & $4.47\times10^9$ & & \\
  $\alpha_{123}$  & $Jm^9/C^6$ & $4.91\times10^9$ & &\\
  \hline
  $c_{11}$ & $J/m^3$ & $27.5\times 10^{10}$ & &\\
  $c_{12}$ & $J/m^3$ & $17.9\times10^{10}$ & & \cite{hlinka2009piezoelectric}\\
  $c_{44}$  & $J/m^3$ & $5.43\times 10^{10}$ & &\\
  \hline
  $q_{11}$ & $Jm/C^2$ & $14.2\times10^{9}$  & &\\
  $q_{12}$ & $Jm/C^2$ & $-0.74\times10^{9}$  & & \cite{hlinka2009piezoelectric}\\
  $q_{44}$ & $Jm/C^2$ & $6.28\times10^{9}$ & &\\
  \hline
  $G_{11}$ & $Jm^3/C^2$ & $51\times10^{-11}$ & &\\
  $G_{12}$ & $Jm^3/C^2$ & $-2\times10^{-11}$ & & \cite{hlinka2009piezoelectric}\\
  $G_{44}$ & $Jm^3/C^2$ & $2\times10^{-11}\times10^{-10}$ & &\\
  \hline
  $\Gamma$ & $C^2/(Jms)$ & $4\times10^4$ & & \cite{hlinka2009piezoelectric}\\
  \hline
  $\varepsilon_B$& $1$   & $10$ & & \cite{tagantsev2010domains} \\
  \hline
   $N$ & $1/m^3$ & $1\times10^{23}$ & $2\times10^{27}$ & \cite{sturman2017charged}\\
  \hline
   $\gamma_n$ & $m^3/s$ & $1.25\times10^{-15}$ & $1\times10^{-17}$ & \\

   $\gamma_p$ & $m^3/s$ & $1.25\times10^{-15}$ & $1\times10^{-17}$ & ** \\
  
   $\gamma$ & $m^3/s$ & $1.25\times10^{-15}$ & $1\times10^{-17}$ & \\
  \hline
   $\mu_n$ & $m^2/(V\cdot s)$ & $1.3\times10^{-5}$ & $1.1\times10^{-6}$ & \cite{yoo2004electronic}\\

   $\mu_p$ & $m^2/(V\cdot s)$ & $0.8\times10^{-5}$ & $1\times10^{-6}$ & \cite{yoo2004electronic}\\
  \hline
   $D_n$ & $m^2/s$ &$3.3\times 10^{-7}$ & $2.8\times 10^{-8}$ &$k_B T\mu_n/e$ \\
 
   $D_p$ & $m^2/s$ & $2\times 10^{-7}$ & $2.55\times 10^{-8*
}$ &$k_B T\mu_p/e$ \\
  \hline
   $B_n$ & $m^3/(C \cdot s)$ &  & -250 & \\
     &  & $-0.053$ &  & *\\
  
   $B_p$ & $m^3/(C \cdot s)$ &  & -200 & \\
  \hline
   $I$ & $J/(m^2 s)$ & $100$ & & ***\\
  \hline
   $ \hslash \omega$ & J & $5.45 \times 10^{-19}$ & & $3.4\times e$ \\
  \hline
   \hline
    $ g $ & $m^{-3}s^{-1}$ & $2.8\times 10^{23}$ & $1\times10^{38}$& $I/\hslash\omega w$  \\
 \hline
 $h $ & $m$ & $0.8\times 10^{-3}$ & $4\times10^{-8}$& sample height \\
 \hline
 $l $ & $m$ & $2.87\times 10^{-3}$ & $2\times10^{-7}$& sample length \\
  \hline
 $w $ & $m$ & $0.65\times 10^{-3}$ & & sample width(depth) \\
 \hline
  $A $ & $m^2$ & $1.86\times 10^{-6}$ & & electrode area  \\
 \hline
 $U $ & $V$ & $300$ & $3$& voltage \\
  \hline
 $e $ & $C$ & $1.6\times 10^{-19}$ & $ $& elementary charge \\
 \hline
 $\tau_{e} $ & $s$ & $1.6\times 10^{-8}$ & $10^{-10}$& electron lifetime $\frac{2}{N \gamma_n}$ \\
 \hline
\botrule
\end{tabular}
\label{tab:par}
\end{table}

\newpage

\subsubsection{Calculations}

The model geometry is a block of a [110]-oriented BaTiO$_3$ crystal with thickness $h=40$ nm and length $l=200$ nm.
Mechanically free boundary conditions were used for all surfaces.
The bottom and top surfaces had the electric potential fixed to $\varphi=0$ and $\varphi=U$, respectively, where $U=3$ V.
The condition of zero free charge was set on the side surfaces.
The initial conditions for polarization were set in the form of periodic stripes of [010]- and [100]-oriented domains, the thickness of the stripes being in agreement with Mitsui and Furuichi theory \cite{mitsui1953domain}.
Initial mechanical displacement corresponds to a stress-free sample.
The model is numerically solved by the finite element method with a time-dependent solver in COMSOL 5.3.
\newpage
\backmatter
\bmhead{Supplementary information}

\setcounter{figure}{0}
\renewcommand{\figurename}{Supplementary Fig.}

\begin{figure}[b]
\centering
\includegraphics[]{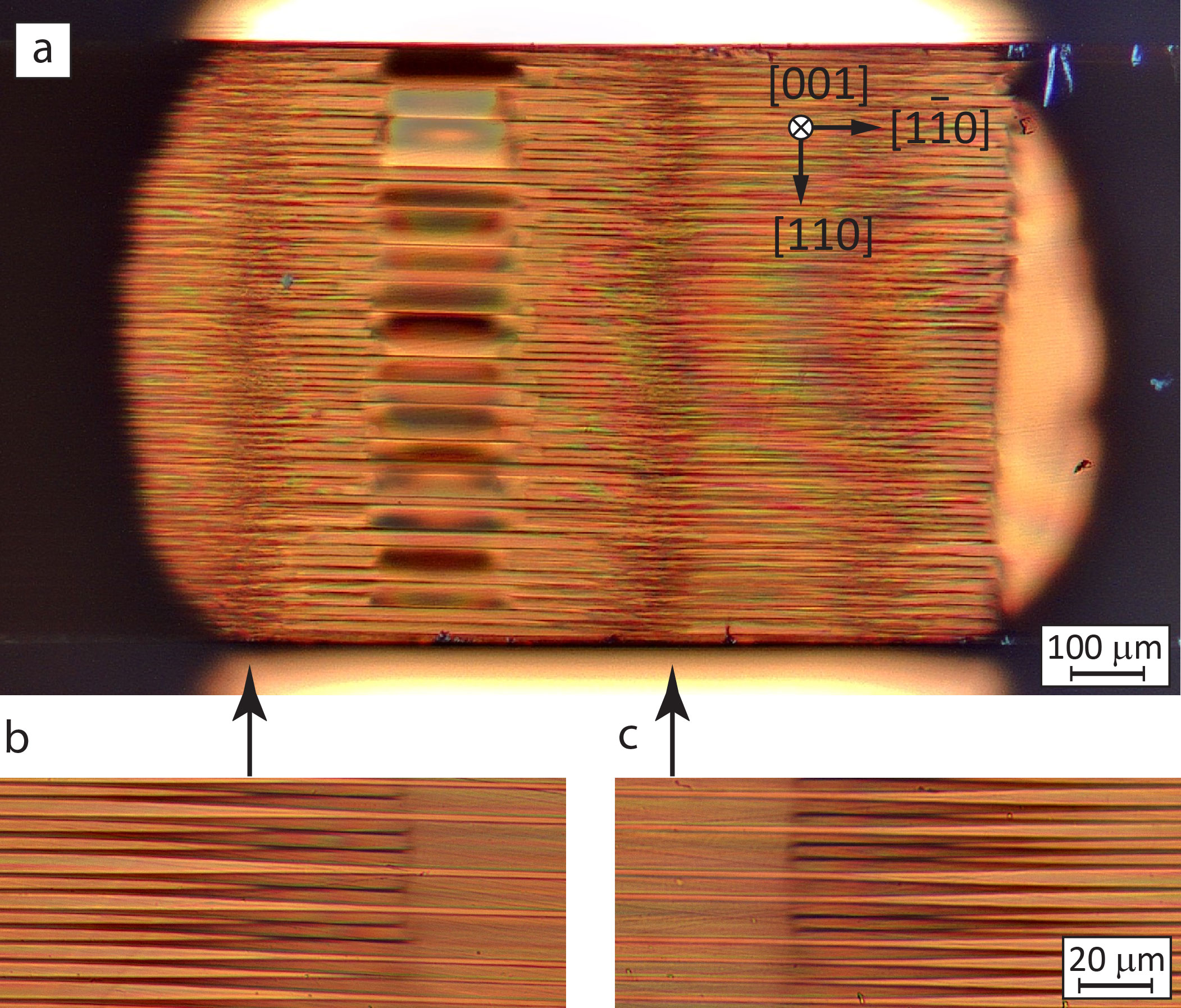}
\caption{
NDWs bending and subsequent fragmentation of thin domains under UV illumination.
Entire sample showing broad dark stripes corresponding to NDW bending (a).
Magnified views highlighting local bending and subsequent domain fragmentation (b–c) .
}
\label{NDWbreaking}
\end{figure}

\begin{figure}[h]
\centering
\includegraphics[width=0.5\textwidth]{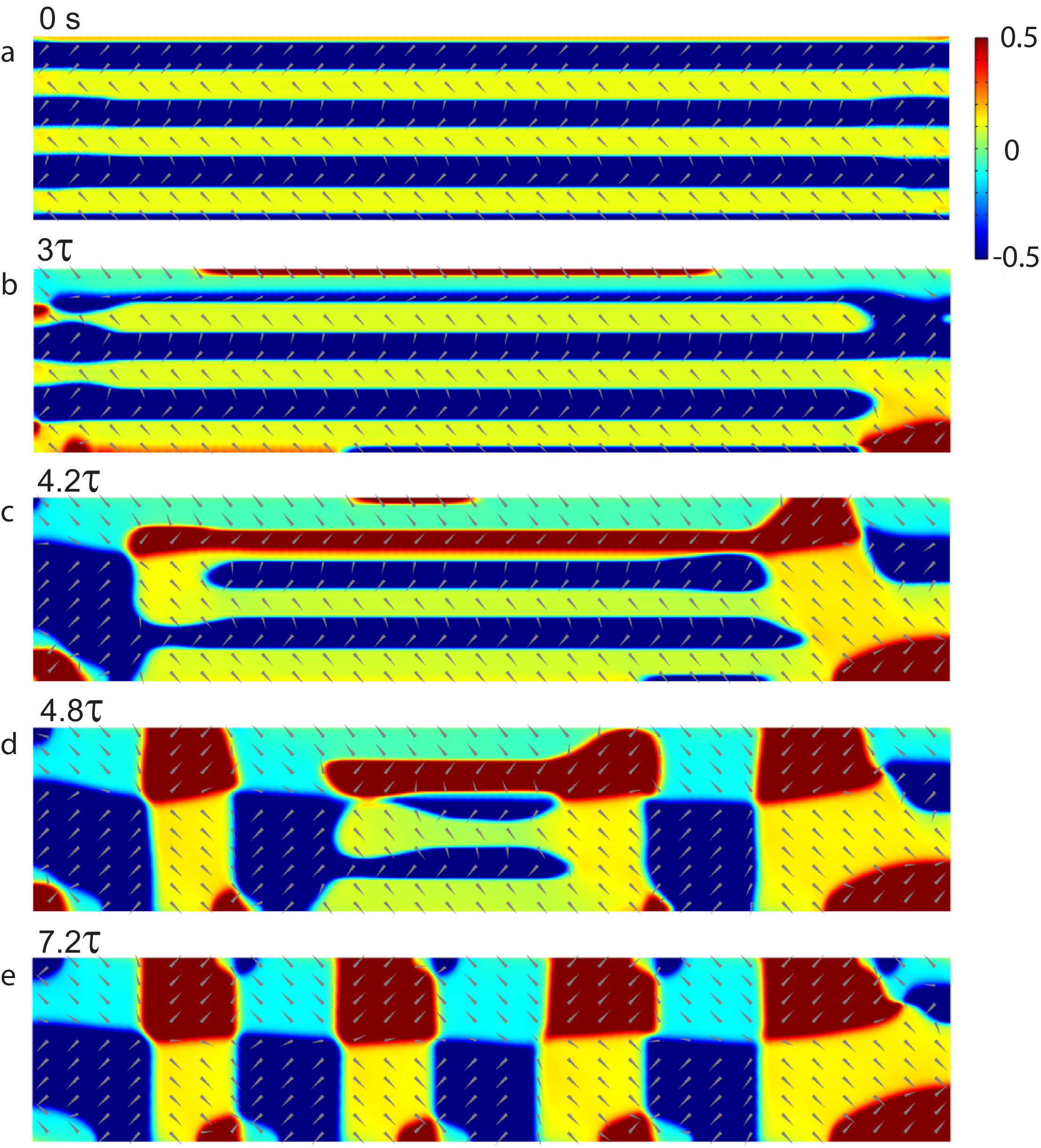}
\caption{Phase-field modeling of domain evolution in zero electric field.
(a) The original system of horizontal neutral domain walls; (b–d) intermediate stages of the transformation; (e) a set of domain walls with a certain degree of disorder to which the simulation converged.
Arrows show polarization; the color map is introduced for the value $1.5 P_x + P_y$.
Units of characterization time $\tau $ are used for the Eq. (\ref{Timescale}) with model parameters.
\label{NoField}}
\end{figure}

\newpage

\bmhead{Relation of the photovoltaic coefficients to the classical ones}\label{tensorbeta}

The classically introduced bulk photovoltaic tensor $\sigma_{ijk}$ relates the current density $J_i$ with the components of the electric field of the incident light $E^{(l)}_k$
 \begin{equation}\label{BPVE1}
J_i=  \sigma_{ijk}E^{(l)}_jE^{(l)}_k,\\
\end{equation}
As the intensity of light is proportional to the square of its electric field, BPVE is often introduced in terms of the light intensity $I$.
For the symmetry of the ferroelectric phase under consideration, the BPVE tensor relationships take the form (see, e.g., ref. \cite{inoue2015giant})
 \begin{eqnarray}\label{BPVEtens}
\label{curr1}J_1=I \cdot 2 \beta_{15}e_1e_3,\\
\label{curr1}J_2=I \cdot  2 \beta_{15}e_2e_3,\\
\label{curr1}J_3=I \cdot  2 \beta_{31}(e_1^2+e_2^2)+\beta_{33}e_3^2,
\end{eqnarray}
where $e_i$ is the light polarization (unit vector), the third axis points along the spontaneous polarization.
We created such illumination conditions that light polarization may be considered random and uniform.
Averaging of Eqs. (\ref{BPVEtens}) over light polarization vectors yields
 \begin{eqnarray}\label{BPVEtensAv}
\label{curr1}J_1=0,\\
\label{curr1}J_2=0,\\
\label{curr1}J_3=I (\frac{2}{3}\beta_{31}+\frac{1}{3}\beta_{33}).
\end{eqnarray}
Our model for BPVE is consistent with Eqs. (\ref{BPVEtensAv}) if the following relationship holds between the moduli. 
 \begin{equation}\label{recalcB}
\frac{2}{3}\beta_{31}+\frac{1}{3}\beta_{33}=\frac{e  P_S}{\hslash\omega w}\left(\frac{B_n}{\gamma_nN_{\circ}}+\frac{B_p}{\gamma_pN_{\bullet}}\right).
\end{equation}
Indeed, in a stationary uniform case with neglected quadratic recombination, our governing equations for the charge balance (\ref{cbal}), (\ref{Absorption}) reduce to
 \begin{equation}
 \label{gammas}
\gamma_nnN_{\circ}=\gamma_ppN_{\bullet}=I/\hslash\omega w.
\end{equation}
Expressing the steady-state concentrations of electrons and holes, and substituting them into equations for their currents, (\ref{currents}), one gets
 \begin{eqnarray}
 \label{22}
j^{(n)}_3=\frac{e B_n I P_S}{\sqrt{2}\hslash\omega w \gamma_nN_{\circ}} \\
j^{(p)}_3=\frac{e B_p I P_S}{\sqrt{2}\hslash\omega w \gamma_pN_{\bullet}},
\end{eqnarray}
which confirms (\ref{recalcB}).

\bmhead{Acknowledgements}

The authors acknowledge the support provided by the Ferroic Multifunctionalities project, supported by the Ministry of Education, Youth, and Sports of the Czech Republic.
Project No. CZ.02.01.01/00/22\underline{\,\,\,}008/0004591, co-funded by the European Union.

\bmhead{Data availability}

The data that support this work are available via Zenodo at https://doi.org/10.5281/zenodo.19998499





\bibliography{sn-bibliography}

\end{document}